\documentclass{article}%
\usepackage{amsmath}
\usepackage{amsfonts}
\usepackage{amssymb}
\usepackage{graphicx}%
\providecommand{\U}[1]{\protect\rule{.1in}{.1in}}

\begin{document}

\title{From Inference to Physics\thanks{Presented at MaxEnt 2008, the 28th
International Workshop on Bayesian Inference and Maximum Entropy Methods in
Science and Engineering (July 8-13, 2008, Boraceia Beach, Sao Paulo, Brazil).}}
\author{Ariel Caticha\\{\small Department of Physics, University at Albany-SUNY, }\\{\small Albany, NY 12222, USA.}}
\date{}
\maketitle

\begin{abstract}
Entropic dynamics, a program that aims at deriving the laws of physics from
standard probabilistic and entropic rules for processing information, is
developed further. We calculate the probability for an arbitrary path followed
by a system as it moves from given initial to final states. For an
appropriately chosen configuration space the path of maximum probability
reproduces Newtonian dynamics.

\end{abstract}

\section{Introduction}

It is not unusual to hear that science consists in using information about the
world for the purpose of predicting, modeling, and/or controlling phenomena of
interest. If this vague image turns out to be even remotely accurate then we
expect that the laws of science should reflect, at least to some extent, the
methods for manipulating information. Here we wish to entertain a far more
radical hypothesis: perhaps \emph{the laws of physics are nothing but rules of
inference}. In this view the laws of physics are not laws of nature but are
merely the rules we follow when processing the information that happens to be
relevant to the physical problem at hand. The evidence supporting this notion
is already quite considerable: most of the formal structure of statistical
mechanics \cite{Jaynes57} and of quantum theory (see e.g. \cite{Caticha98})
can be derived as examples of inference.

The basic difficulty is that the available information is usually incomplete
and one must learn to handle uncertainty. This requires addressing three
problems; the first two have been satisfactorily solved, the third one has
not. First, one must represent one's partial state of knowledge as a web of
interconnected beliefs with no internal inconsistencies; the tools to do it
are probabilities \cite{Cox46, Jaynes03}. Second, when new information becomes
available the beliefs must be correspondingly updated. The instrument for
updating is relative entropy and the resulting procedure---the ME\ method---is
the only candidate that can claim universal applicability. The ME method is
based on the recognition that prior information is valuable and should not be
revised except when demanded by new evidence; it can handle arbitrary priors
and arbitrary constraints; it includes MaxEnt and Bayes' rule as special
cases; and it provides a quantitative assessment of the extent that
distributions that deviate from the entropy maximum are ruled out. (See e.g.
\cite{Caticha07, Caticha08}.)

The third problem is trickier. When we say that \textquotedblleft the laws of
physics are not laws of nature\textquotedblright\ we do not mean that physics
can be derived without any input from nature; quite the opposite. The
statement \textquotedblleft physics is inference\textquotedblright\ comes with
considerable fine print. It implicitly assumes that one is doing inference
about the \textquotedblleft right things\textquotedblright\ on the basis of
the \textquotedblleft right information.\textquotedblright\ The third and so
far unsolved problem is that of identifying the questions that are interesting
and the information that is relevant about a particular physical
situation---this is where the connection to nature lies. The current
approaches cannot be called a method---ultimately there is no scientific
\textquotedblleft method.\textquotedblright\ We have learned from
experience---a euphemism for trial and error, mostly error---which pieces of
information happen to work well in each specific situation. Recent results,
however, in model selection \cite{Rodriguez05} and in the development of a
quantitative theory of inquiry and of relevance \cite{Knuth05} represent
considerable progress and point the way towards more systematic approaches.

In any case, once the relevant information has been identified, if the laws of
physics are merely rules of inference, then we should be able to derive them.
Our main concern is to derive laws of dynamics and the challenge---of
course---is to avoid assuming the very laws of motion that we set out to
derive. The formalism, which we refer to as \emph{entropic dynamics}
\cite{Caticha01, Caticha0305}, is of general applicability but to be specific
we focus on the example of particle mechanics.

In a previous paper \cite{CatichaCafaro07} we derived Newtonian mechanics
without assuming a principle of least action, or concepts of force, or
momentum, or mass, and not even the notion of an absolute Newtonian time. None
of these familiar concepts are part of the input to the theory---they are all
derived. As described in \cite{CatichaCafaro07} the crucial step was the
selection of a suitable statistical model for the configuration space of a
system of particles which amounts to specifying both the subject matter and
the relevant background information.

The objective of the present paper is to develop the formalism of entropic
dynamics further. We address the same dynamically interesting question: Given
an initial and a final state, what trajectory will the system follow? In
\cite{CatichaCafaro07} we had calculated the path of maximum probability and
we showed that it corresponds to Newtonian dynamics. But the available
information does not single out a unique path; here---and this is our main
result---we calculate the probability for any arbitrary path between the given
initial and final states. As a first application we verify that indeed the
most probable path reproduces our earlier result. A more detailed study of
fluctuations and diffusion about the Newtonian path will, however, be left for
a future publication.

We conclude with brief remarks about the asymmetry between past and future as
seen from the unfamiliar perspective of entropic dynamics, and about a
possible connection between this work and Nelson's derivation of quantum
mechanics as a peculiar kind of diffusion process \cite{Nelson6685}.

\section{Physical space and configuration space}

Consider one particle (or many) living in our familiar \textquotedblleft
physical\textquotedblright\ space (whatever this might ultimately mean). There
is a useful distinction to be drawn between this physical\ space $\mathcal{Y}$
and the space of states or configuration space $\mathcal{X}$. For simplicity
we will assume that physical space $\mathcal{Y}$ is flat and three
dimensional; its geometry is given by the Euclidean metric $ds^{2}=\delta
_{ab}dy^{a}dy^{b}$---generalizations are straightforward. The configuration
space $\mathcal{X}$ for a single particle will also be assumed to be three
dimensional but it need not be flat. The interesting dynamics will arise from
its curvature. The main additional ingredient is that there is an irreducible
uncertainty in the location of the particle. Thus, when we say that the
particle is at the point $x\in\mathcal{X}$ what we mean is that its
\textquotedblleft physical\textquotedblright, \textquotedblleft
true\textquotedblright\ position $y\in\mathcal{Y}$ is somewhere in the
vicinity of $x$. This leads us to associate a probability distribution
$p(y|x)$ to each point $x$ and the space $\mathcal{X}$ is thus transformed
into a statistical manifold: a point $x$ is not a structureless dot but a
fuzzy probability distribution. The origin of these uncertainties is, at this
point, left unspecified.

In \cite{CatichaCafaro07} we adopted a Gaussian model,
\begin{equation}
p(y|x)=\frac{\gamma^{1/2}(x)}{(2\pi)^{3/2}}\,\exp\left[  -\frac{1}{2}%
\gamma_{ab}(x)(y^{a}-x^{a})(y^{b}-x^{b})\right]  ,\label{Gaussian}%
\end{equation}
where $\gamma=\det\gamma_{ab}$. It incorporates the physically relevant
information of an estimate of the particle position, $\langle y^{a}%
\rangle=x^{a}$, and of its small uncertainty as given by the covariance
matrix,
\begin{equation}
\tilde{\gamma}^{ab}=\langle(y^{a}-x^{a})(y^{b}-x^{b})\rangle~,
\end{equation}
which is the inverse of $\gamma_{ab}$, $\tilde{\gamma}^{ab}\gamma_{bc}%
=\delta_{c}^{a}$. The choice of Gaussians is physically plausible but not
strictly necessary. We are trying to predict behavior at macro-scales in terms
of assumptions we make (that is, information we assume) about what is going on
at some intermediate meso-scales which themselves are the result of happenings
at still shorter micro-scales about which we know absolutely nothing. If the
fuzziness in position that we postulate at the meso-scale is the result of
many unknown microscopic influences going on at a much smaller micro-scale
then general arguments such as the central limit theorem lead us to expect
Gaussians as the plausible mesoscopic distributions for a very wide variety of
microscopic conditions.

To conclude the specification of the model we further impose that the
Gaussians be spherically symmetric\footnote{This corresponds to the Newtonian
assumption that space is locally isotropic.} with a small but non-uniform
variance $\sigma^{2}(x)$ conveniently expressed in terms of a small constant
$\sigma_{0}^{2}$ modulated by a (positive) scalar field $\Phi(x)$,
\begin{equation}
\gamma_{ab}(x)=\frac{1}{\sigma^{2}(x)}\delta_{ab}=\frac{\Phi(x)}{\sigma
_{0}^{2}}\delta_{ab}~.\label{gamma}%
\end{equation}

The next feature is automatic, it requires no further assumptions: the
configuration space $\mathcal{X}$, when viewed as a statistical manifold,
inherits a geometry from the distributions $p(y|x)$. The distance between two
neighboring distributions $p(y|x)$ and $p(y|x+dx)$ is the \emph{unique}
measure of the extent that one distribution can be statistically distinguished
from the other---\emph{distinguishability is distance }(and possibly
vice-versa, but that is a story for another paper \cite{Caticha0305}). It is
given by the information metric of Fisher and Rao \cite{Amari00, Caticha08},
\begin{equation}
d\ell^{2}=g_{ab}\,dx^{a}dx^{b}\quad\text{with}\quad g_{ab}=\int
dy\,p(y|x)\frac{\partial\log p(y|x)}{\partial x^{a}}\frac{\partial\log
p(y|x)}{\partial x^{b}}~.\label{info metric a}%
\end{equation}
The corresponding volume element is $dv=g^{1/2}(x)d^{3}x$ where $g=\det
g_{ab}$. Substituting (\ref{Gaussian}) and (\ref{gamma}) into
(\ref{info metric a}) we obtain the information metric for the manifold of
spherically symmetric Gaussians,
\begin{equation}
g_{ab}(x)=\frac{1}{\sigma^{2}(x)}\left(  \delta_{ab}+6\partial_{a}%
\sigma\partial_{b}\sigma\right)  \approx\frac{\Phi(x)}{\sigma_{0}^{2}}%
\delta_{ab}=\gamma_{ab}(x)~,\label{info metric b}%
\end{equation}
provided $\sigma_{0}$ is sufficiently small.

\section{The probability of a path}

The fact that the state $x$ of the particle might be unknown is described by a
distribution $P(x)$. The path of a particle is an ordered sequence of $N+1$
positions $\{x_{0}\ldots x_{N}\}$. We want to calculate the probability
\begin{equation}
P(x_{1}\ldots x_{N-1}|x_{0}x_{N})dx_{1}\ldots dx_{N-1}~\label{Path Prob}%
\end{equation}
that the path passes through small volume elements $dx_{n}$ at the
intermediate points $x_{1}\ldots x_{N-1}$. Since
\begin{equation}
P(x_{1}\ldots x_{N-1}|x_{0}x_{N})=\frac{P(x_{1}\ldots x_{N-1}x_{N}|x_{0}%
)}{P(x_{N}|x_{0})}~,\label{Path Prob Cond}%
\end{equation}
our immediate interest will be to assign a probability $P(x_{1}\ldots
x_{N}|x_{0})$ of the ordered path $\{x_{1}\ldots x_{N}\}$ starting at $x_{0}$.

Note that an external time has not been introduced. It is true that the path
is ordered so that \emph{along a given path} the point $x_{n}$ reached after
$n$ steps could be construed to occur \emph{later }than the point reached at
the \emph{previous} step, $x_{n-1}$. But most important elements implicit in
the notion of time are conspicuously absent. For example, we still have no way
to order temporally the point $x_{n}$ reached along one path with the point
$x_{n^{\prime}}^{\prime}$ reached along a different path. Statements to the
effect that one occurs earlier or later or simultaneously with the other are,
at this point, completely meaningless. We have not introduced a notion of
\emph{simultaneity} and therefore we do not have a notion of an \emph{instant
of time}. Furthermore we do not have a notion of \emph{duration }either; we
have not introduced a way to compare or measure intervals for the successive
steps. The statement that a certain step took, say, twice as long as the
previous step is, at this point, meaningless.

Without the notions of instant or of interval we do not have time. An
important part of the program of deriving physics from inference consists in
understanding how and where these temporal concepts arise.

\subsection{The single-step probability}

To warm up we first calculate the probability $P(x_{1}|x_{0})$ to reach
$x_{1}$ in a single step. Since we are ignorant not only about the
true\ position $y_{1}$ but also about the configuration space position $x_{1}$
the relevant distribution of interest is the joint distribution $P_{J}%
(x_{1}y_{1}|x_{0})$. We shall choose the distribution $P_{J}(x_{1}y_{1}%
|x_{0})$ using the ME method, that is, by maximizing the single-step entropy
\cite{Caticha08}
\begin{equation}
\mathcal{S}_{1}[P_{J},P_{J}^{\prime}]=-%
{\textstyle\int}
dx_{1}dy_{1}\,P_{J}(x_{1}y_{1}|x_{0})\log\frac{P_{J}(x_{1}y_{1}|x_{0})}%
{P_{J}^{\prime}(x_{1}y_{1}|x_{0})}~.\label{Sigma a}%
\end{equation}

\noindent\textbf{The prior:} $P_{J}^{\prime}(x_{1}y_{1}|x_{0})$ represents
partial knowledge about the variables $x_{1}$ and $y_{1}$ \emph{before} we
incorporate any information in the form of constraints. Let
\begin{equation}
P_{J}^{\prime}(x_{1}y_{1}|x_{0})=P^{\prime}(x_{1}|x_{0})P^{\prime}(y_{1}%
|x_{0}x_{1})~,
\end{equation}
and focus first on $P^{\prime}(x_{1}|x_{0})$. At this point it is not yet
known how $x_{1}$ is related to $x_{0}$ or to $y_{1}$. We do know that
$x_{1}\in\mathcal{X}$ labels some probability distribution $p(y|x_{1})$ in
$\mathcal{X}$, eqs.(\ref{Gaussian}, \ref{gamma}), but we do not yet know that
it is the distribution of $y_{1}$, $p(y_{1}|x_{1})$. Thus, we are maximally
ignorant about $x_{1}$ and, accordingly, we choose a uniform distribution
$P^{\prime}(x_{1}|x_{0})\propto g^{1/2}(x_{1})$. For the second factor
$P^{\prime}(y_{1}|x_{0}x_{1})$ we argue that the variables $y_{1}$ are meant
to represent the actual (uncertain) coordinates of a particle; we assume that
in the absence of any information to the contrary the distribution of $y_{1}$
remains unchanged from the previous step, $P^{\prime}(y_{1}|x_{0}%
x_{1})=p(y_{1}|x_{0})$. Thus, the joint prior $P^{\prime}$ is
\begin{equation}
P_{J}^{\prime}(x_{1}y_{1}|x_{0})\propto g^{1/2}(x_{1})p(y_{1}|x_{0}%
)~.\label{PJ a}%
\end{equation}

\noindent\textbf{The constraint:} Next we incorporate the piece of information
that establishes the relation between $x_{1}$ and $y_{1}$. This is the
constraint that demands updating from the prior to the posterior. The
posterior $P_{J}(x_{1}y_{1}|x_{0})$ belongs to the family of distributions
\begin{equation}
P_{J}(x_{1}y_{1}|x_{0})=P(x_{1}|x_{0})P(y_{1}|x_{1}x_{0})\label{PJ b}%
\end{equation}
where $P(x_{1}|x_{0})$ is arbitrary and the second factor is constrained to be
of the form $P(y_{1}|x_{0}x_{1})=p(y_{1}|x_{1})$.\footnote{Incidentally, this
is an example of an application of the ME method where the constraints are not
in the form of expected values; these are not \textquotedblleft
linear\textquotedblright\ constraints.}

Substituting (\ref{PJ a}) and (\ref{PJ b}) into (\ref{Sigma a}) and
rearranging gives
\begin{equation}
\mathcal{S}_{1}[P_{J},P_{J}^{\prime}]=-%
{\textstyle\int}
dx_{1}\,P(x_{1}|x_{0})\left[  \log\frac{P(x_{1}|x_{0})}{g^{1/2}(x_{1}%
)}-S(x_{1},x_{0})\right]  ~,\label{Sigma b}%
\end{equation}
where
\begin{equation}
S(x_{1},x_{0})=-%
{\textstyle\int}
dy_{1}\,p(y_{1}|x_{1})\log\frac{p(y_{1}|x_{1})}{p(y_{1}|x_{0})}~.
\end{equation}
To determine $P(x_{1}|x_{0})$ maximize eq.(\ref{Sigma b}) subject to
normalization. The first term in eq.(\ref{Sigma b}) makes $x_{1}$ as random as
possible; by itself it would lead to a uniform distribution $P(x_{1}%
|x_{0})\propto g^{1/2}(x_{1})$. The second term in eq.(\ref{Sigma b}) brings
$x_{1}$ as close as possible to $x_{0}$; it would make $P(x_{1}|x_{0}%
)\propto\delta(x_{1}-x_{0})$ and push $S(x_{1},x_{0})$ towards its maximum
value, $S(x_{0},x_{0})=0$. The compromise between these two opposing
tendencies is
\begin{equation}
P_{1}(x_{1}|x_{0})=\frac{1}{z(x_{0})}g^{1/2}(x_{1})e^{S(x_{1},x_{0}%
)}~,\label{single step Prob}%
\end{equation}
where $z(x_{0})$ is an appropriate normalization constant.

The probability (\ref{single step Prob}) represents a \emph{discontinuous}
jump from $x_{0}$ to $x_{1}$ in a single step. No information has been imposed
to the effect that the particle \textquotedblleft moves\textquotedblright%
\ from $x_{0}$ to $x_{1}$ along a continuous trajectory. This is done next by
assuming that the continuous trajectory can be approximated by a sequence of
$N$ steps where $N$ is large.

\subsection{The $N$-step probability}

To assign the probability $P(x_{1}\ldots x_{N}|x_{0})$ for a path we focus on
the joint distribution $P_{J}(x_{1}y_{1}\ldots x_{N}y_{N}|x_{0})\,$and choose
the distribution that maximizes
\begin{equation}
\mathcal{S}_{N}[P_{J},P_{J}^{\prime}]=-%
{\textstyle\int}
(%
{\textstyle\prod\limits_{n=1}^{N}}
dx_{n}dy_{n})\,P_{J}\log\frac{P_{J}}{P_{J}^{\prime}}~.\label{Sigma c}%
\end{equation}

\noindent\textbf{The prior:} To assign $P_{J}^{\prime}(x_{1}y_{1}\ldots
x_{N}y_{N}|x_{0})$ consider the path in the spaces of $x$s and $y$s separately
($x\in$ $\mathcal{X}$ and $y\in\mathcal{Y}$).
\begin{equation}
P_{J}^{\prime}(x_{1}y_{1}\ldots x_{N}y_{N}|x_{0})=P^{\prime}(x_{1}\ldots
x_{N}|x_{0})P^{\prime}(y_{1}\ldots y_{N}|x_{0}x_{1}\ldots x_{N}%
)~.\label{P prior a}%
\end{equation}
The first factor is the prior probability of a path in the space
$\mathcal{X}^{N}$. To the extent that we know nothing about the relation
between successive $x$s we choose a uniform distribution in the space of
paths,
\begin{equation}
P^{\prime}(x_{1}\ldots x_{N}|x_{0})\propto%
{\textstyle\prod\limits_{n=1}^{N}}
g^{1/2}(x_{n})\,.\label{P prior b}%
\end{equation}
The second factor is the prior probability of a path in the space
$\mathcal{Y}^{N}$. We assume that in the absence of any information to the
contrary the distribution of the $n$-th step $y_{n}$ retains memory only of
the immediately preceding $x_{n-1}$,%

\begin{equation}
P^{\prime}(y_{1}\ldots y_{N}|x_{0}x_{1}\ldots x_{N})=%
{\textstyle\prod\limits_{n=1}^{N}}
p(y_{n}|x_{n-1})~.\label{P prior c}%
\end{equation}
(This is not quite a Markov process; the distribution of $y_{n}$ does not
retain memory of the immediately preceding $y_{n-1}\in\mathcal{Y}$, only of
$x_{n-1}\in\mathcal{X}$.)

\noindent\textbf{The constraint:} Next we impose the information that relates
$y_{n}$ to its corresponding $x_{n}$. The posterior
\begin{equation}
P_{J}(x_{1}y_{1}\ldots x_{N}y_{N}|x_{0})=P(x_{1}\ldots x_{N}|x_{0}%
)P(y_{1}\ldots y_{N}|x_{0}x_{1}\ldots x_{N})
\end{equation}
is constrained to belong to the family of distributions such that
\begin{equation}
P(y_{1}\ldots y_{N}|x_{0}x_{1}\ldots x_{N})=%
{\textstyle\prod\limits_{n=1}^{N}}
p(y_{n}|x_{n})~.\label{constraint}%
\end{equation}
Substituting $P_{J}$ and $P_{J}^{\prime}$ into eq.(\ref{Sigma c}) and
rearranging gives
\begin{align}
\mathcal{S}_{N}[P_{J},P_{J}^{\prime}]  & =-%
{\textstyle\int}
(%
{\textstyle\prod\limits_{n=1}^{N}}
dx_{n})\,P(x_{1}\ldots x_{N}|x_{0})\log\frac{P(x_{1}\ldots x_{N}|x_{0})}{%
{\textstyle\prod\limits_{n=1}^{N}}
g^{1/2}(x_{n})}\nonumber\\
& +%
{\textstyle\int}
(%
{\textstyle\prod\limits_{n=1}^{N}}
dx_{n})\,P(x_{1}\ldots x_{N}|x_{0})%
{\textstyle\sum\limits_{n=1}^{N}}
S(x_{n},x_{n-1})\label{Sigma d}%
\end{align}
where
\begin{equation}
S(x_{n},x_{n-1})=-%
{\textstyle\int}
dy_{n}\,p(y_{n}|x_{n})\log\frac{p(y_{n}|x_{n})}{p(y_{n}|x_{n-1})}%
\end{equation}
As before the two integrals in eq.(\ref{Sigma d}) represent opposing
tendencies. The first integral seeks to make $P(x_{1}\ldots x_{N}|x_{0})$ as
random as possible with $x_{n}$ completely uncorrelated to $x_{n-1}$. The
second integral introduces strong correlations; it brings $x_{n}$ as close as
possible to the preceding $x_{n-1}$.

\noindent\textbf{The main result:} Varying $P(x_{1}\ldots x_{N}|x_{0})$ to
maximize eq.(\ref{Sigma d}) subject to normalization gives the probability
density for a path starting at the initial position $x_{0}$,
\begin{equation}
P_{N}(x_{1}\ldots x_{N}|x_{0})=\frac{1}{Z(x_{0})}[%
{\textstyle\prod\limits_{n=1}^{N}}
g^{1/2}(x_{n})]\exp[%
{\textstyle\sum\limits_{n=1}^{N}}
S(x_{n},x_{n-1})]\label{Path Prob a}%
\end{equation}
where $Z(x_{0})$ is the appropriate normalization constant.

The probability density for the $N$-step path between given initial and final
positions $x_{0}$ and $x_{N}$ is given by (\ref{Path Prob Cond}) where
$P_{N}(x_{N}|x_{0})$ is obtained from (\ref{Path Prob a}),
\begin{equation}
P_{N}(x_{N}|x_{0})=\frac{g^{1/2}(x_{N})}{Z(x_{0})}%
{\textstyle\int}
[%
{\textstyle\prod\limits_{n=1}^{N-1}}
dx_{n}g^{1/2}(x_{n})]\exp[%
{\textstyle\sum\limits_{n=1}^{N}}
S(x_{n},x_{n-1})]~.\label{Path Prob b}%
\end{equation}
Substituting back into (\ref{Path Prob Cond}) gives the desired answer
\begin{equation}
P_{N}(x_{1}\ldots x_{N-1}|x_{0}x_{N})=\frac{1}{Z(x_{0},x_{N})}[%
{\textstyle\prod\limits_{n=1}^{N-1}}
g^{1/2}(x_{n})]\exp[%
{\textstyle\sum\limits_{n=1}^{N}}
S(x_{n},x_{n-1})]\,,\label{Path Prob c}%
\end{equation}
where $Z(x_{0},x_{N})$ is the appropriate normalization. Equations
(\ref{Path Prob a}) and (\ref{Path Prob c}) are the main results of this paper.

\section{The most probable path}

We restrict our analysis of eq.(\ref{Path Prob c}) to calculating the most
probable path from the initial position $x_{0}$ to the final $x_{N}$. For
fixed volume elements, $dV_{n}=g^{1/2}(x_{n})dx_{n}=dV$, the path of maximum
probability is that which maximizes
\begin{equation}
A(x_{1}\ldots x_{N-1}|x_{0}x_{N})=%
{\textstyle\sum\limits_{n=1}^{N}}
S(x_{n},x_{n-1})\,,
\end{equation}
where $x_{0}$ and $x_{N}$ are fixed. The maximum probability path is the
polygonal path that brings the successive $x$s as \textquotedblleft
close\textquotedblright\ to each other as possible. For large $N$ we expect
this to be the shortest path between the given end points and, as shown below,
this is indeed the case. The variation of $A$ is
\begin{align}
\delta A  & =%
{\textstyle\sum\limits_{n=1}^{N-1}}
\frac{\partial}{\partial x_{n}^{c}}[%
{\textstyle\sum\limits_{m=1}^{N}}
S(x_{m},x_{m-1})]\delta x_{n}^{c}\nonumber\\
& =%
{\textstyle\sum\limits_{n=1}^{N-1}}
\frac{\partial}{\partial x_{n}^{c}}\left[  S(x_{n},x_{n-1})+S(x_{n+1}%
,x_{n})\right]  \delta x_{n}^{c}~.
\end{align}
For large $N$ we assume that successive $x$s along the path are sufficiently
close together that we can approximate
\begin{equation}
S(x_{n},x_{n-1})=-\frac{1}{2}d\ell_{n,n-1}^{2}=-\frac{1}{2}g_{ab}%
(x_{n-1})(x_{n}^{a}-x_{n-1}^{a})(x_{n}^{b}-x_{n-1}^{b})~.
\end{equation}
Next, introduce a parameter $\lambda$ along the trajectory, $x=x(\lambda)$.
The corresponding velocities $\dot{x}$ are
\begin{equation}
\dot{x}_{n+1/2}\overset{\operatorname{def}}{=}\frac{x_{n+1}-x_{n}}%
{\Delta\lambda},\quad\dot{x}_{n-1/2}\overset{\operatorname{def}}{=}\frac
{x_{n}-x_{n-1}}{\Delta\lambda}\quad\text{and}\quad\dot{x}_{n}\overset
{\operatorname{def}}{=}\frac{x_{n+1}-x_{n-1}}{2\Delta\lambda}.
\end{equation}
Expand,\footnote{We use the standard notation $g_{ac,d}=\partial
g_{ac}/\partial x^{d}$.}
\begin{equation}
g_{ac}(x_{n-1})=g_{ac}(x_{n})-g_{ac,d}(x_{n})\dot{x}_{n-1/2}^{d}\Delta
\lambda+\ldots,
\end{equation}
and rearrange to get
\begin{align}
~\delta A  & =-%
{\textstyle\sum\limits_{n=1}^{N-1}}
\Delta\lambda^{2}[\frac{1}{2}g_{ab,c}(x_{n})\dot{x}_{n+1/2}^{a}\dot{x}%
_{n+1/2}^{b}\nonumber\\
& +g_{ac}(x_{n})\frac{\dot{x}_{n-1/2}^{a}-\dot{x}_{n+1/2}^{a}}{\Delta\lambda
}-g_{ac,d}(x_{n})\dot{x}_{n-1/2}^{d}\dot{x}_{n-1/2}^{a}]~,
\end{align}
where we recognize the acceleration
\begin{equation}
\ddot{x}_{n}\overset{\operatorname{def}}{=}\frac{\dot{x}_{n+1/2}-\dot
{x}_{n-1/2}}{\Delta\lambda}=\frac{x_{n+1}-2x_{n}+x_{n-1}}{(\Delta\lambda)^{2}%
}~.
\end{equation}
Substituting gives
\begin{align}
\delta A  & =\Delta\lambda^{2}%
{\textstyle\sum\limits_{n=1}^{N-1}}
[g_{ac}(x_{n})\ddot{x}_{n}^{a}+g_{ac,d}\left(  \dot{x}_{n}^{a}-\ddot{x}%
_{n}^{a}\frac{\Delta\lambda}{2}\right)  \left(  \dot{x}_{n}^{d}-\ddot{x}%
_{n}^{d}\frac{\Delta\lambda}{2}\right)  \nonumber\\
& -\frac{1}{2}g_{ab,c}\left(  \dot{x}_{n}^{a}+\ddot{x}_{n}^{a}\frac
{\Delta\lambda}{2}\right)  \left(  \dot{x}_{n}^{b}+\ddot{x}_{n}^{b}%
\frac{\Delta\lambda}{2}\right)  ]\delta x_{n}^{c}~.
\end{align}
If the distribution of points along the trajectory is sufficiently dense,
$\Delta\lambda\rightarrow0$, the leading term is
\begin{equation}
\delta A=\Delta\lambda^{2}%
{\textstyle\sum\limits_{n=1}^{N-1}}
[g_{ac}(x_{n})\ddot{x}_{n}^{a}+\frac{1}{2}\left(  g_{ca,b}+g_{cb,a}%
-g_{ab,c}\right)  \dot{x}_{n}^{a}\dot{x}_{n}^{b}]\delta x_{n}^{c}~,
\end{equation}
or
\begin{equation}
\delta A=\Delta\lambda^{2}%
{\textstyle\sum\limits_{n=1}^{N-1}}
g_{ad}(x_{n})[\ddot{x}_{n}^{a}+\Gamma_{bc}^{a}\dot{x}_{n}^{b}\dot{x}_{n}%
^{c}]\delta x_{n}^{d}~,\label{delta A}%
\end{equation}
where $\Gamma_{ab}^{c}$ are Christoffel symbols,
\begin{equation}
\Gamma_{ab}^{c}=\frac{1}{2}g^{cd}\left(  g_{da,b}+g_{db,a}-g_{ab,d}\right)
~.\label{Christoffel a}%
\end{equation}
Setting $\delta A=0$ for arbitrary variations $\delta x_{n}^{d}$ leads to the
geodesic equation,
\begin{equation}
\ddot{x}_{n}^{a}+\Gamma_{bc}^{a}\dot{x}_{n}^{b}\dot{x}_{n}^{c}=0~,
\end{equation}
Incidentally, $\lambda$ turns out to be an affine parameter, that is, up to an
unimportant scale factor it measures the length along the path,
\begin{equation}
d\lambda^{2}=Cg_{ab}dx^{a}dx^{b}~.
\end{equation}
\textbf{Conclusion:} The most probable continuous path between two given end
points is the geodesic that joins them.

\noindent\textbf{Remark:} It is interesting that although the ME inference
implicitly assumed a directionality from the initial $x_{0}$ to the final
$x_{N}$ through the prior for $y_{n}$ which establishes a connection with the
\textquotedblleft previous\textquotedblright\ instant, $p(y_{n}|x_{n-1})$, in
the continuum limit the sense of direction is lost. The most probable
trajectory is fully reversible.

The treatment above is general; it is valid for dynamics on any statistical
manifold. Now we restrict ourselves to the manifold of spherically symmetric
Gaussians defined by (\ref{Gaussian}, \ref{gamma}). The parametrization in
terms of $\lambda$ is convenient but completely arbitrary. Let us instead
introduce a new non-affine \textquotedblleft time\textquotedblright\ parameter
$t=t(\lambda)$ defined by
\begin{equation}
dt=\frac{d\lambda}{2^{1/2}\Phi}\quad\text{or}\quad T_{t}\overset{\text{def}%
}{=}\frac{1}{2\sigma_{0}^{2}}\delta_{ab}\frac{dx^{a}}{dt}\frac{dx^{b}}%
{dt}=\Phi\label{newt time}%
\end{equation}
then the geodesic equation becomes
\begin{equation}
\frac{d^{2}x^{c}}{dt^{2}}+\Gamma_{ab}^{c}\frac{dx^{a}}{dt}\frac{dx^{b}}%
{dt}=-\frac{d^{2}t/d\lambda^{2}}{(dt/d\lambda)^{2}}\frac{dx^{c}}{dt}~,
\end{equation}
and using (\ref{Christoffel a})%

\begin{equation}
\Gamma_{ab}^{c}=\frac{1}{2\Phi}\left(  \partial_{a}\Phi\delta_{b}^{c}%
+\partial_{b}\Phi\delta_{a}^{c}-\partial_{c}\Phi\delta_{ab}\right)
~,\label{Christoffel b}%
\end{equation}
we get
\begin{equation}
\frac{d^{2}x^{a}}{dt^{2}}=\frac{\partial_{a}\Phi}{\Phi}\frac{1}{2}\delta
_{bc}\frac{dx^{b}}{dt}\frac{dx^{c}}{dt}~,
\end{equation}
which, using $T_{t}=\Phi$ from eq.(\ref{newt time}), gives
\begin{equation}
\frac{1}{\sigma_{0}^{2}}\frac{d^{2}x^{a}}{dt^{2}}=\partial_{a}\Phi
\,\frac{T_{t}}{\Phi}=\partial_{a}\Phi~.
\end{equation}
This is Newton's equation. To make it explicit just change notation and call
\begin{equation}
\frac{1}{\sigma_{0}^{2}}\overset{\text{def}}{=}m\quad\text{and}\quad
\Phi(x)\overset{\text{def}}{=}E-V(x)
\end{equation}
where $E$ is a constant. The result is Newton's $F=ma$ and energy
conservation,
\begin{equation}
m\frac{d^{2}x^{a}}{dt^{2}}=-\frac{\partial V(x)}{\partial x^{a}}%
\quad\text{and}\quad\frac{m}{2}\delta_{ab}\frac{dx^{a}}{dt}\frac{dx^{b}}%
{dt}+V(x)=E~.
\end{equation}
\textbf{Conclusion:} We have reproduced the results obtained in
\cite{CatichaCafaro07}. The Newtonian mass $m$ and force $F^{a}=-\partial
_{a}V$ are \textquotedblleft explained\textquotedblright\ in terms of position
uncertainties; the uniform uncertainty $\sigma_{0}$ explains mass, while the
modulating field $\Phi(x)$ explains forces.

The extension to more particles interacting among themselves is
straightforward---see \cite{CatichaCafaro07}. Further analysis will be pursued
elsewhere. Here we only mention that a most remarkable feature of the time $t$
selected according to (\ref{newt time}) is that \emph{isolated} subsystems all
keep the same common time which confirms $t$ as the universal Newtonian time.
Thus, the advantage of the Newtonian time goes beyond the fact that it
simplifies the equations of motion. It is the only choice of time such that
isolated clocks will keep synchronized.

\section{Final remarks}

We conclude with two comments. The first concerns the arrow of time, an
interesting puzzle that has plagued physics ever since Boltzmann \cite{Price96
Zeh01}. The problem is that the laws of physics are symmetric under time
reversal---forget, for the moment, the tiny T violations in K-meson
decay---but everything else in nature seems to indicate a clear asymmetry
between the past and the future. How can we derive an arrow of time from
underlying laws of nature that are symmetric? The short answer is: we can't.

In a few brief lines we cannot do full justice to this problem but we can hint
that entropic dynamics offers a promising new way to address it. We note,
first, that entropic dynamics does not assume any underlying laws of
nature---whether they be symmetric or not. And second, that information about
the past is treated differently from information about the future. Entropic
dynamics does not attempt to explain the asymmetry between past and future.
The asymmetry is accepted as prior information. It is the known but unproven
truth that provides the foundation from which all sorts of other inferences
will be derived. From the point of view of entropic dynamics the problem is
not to explain the arrow of time, but rather to explain the reversibility of
the laws of physics. And in this endeavor entropic dynamics succeeds. Laws of
physics such as $\vec{F}=m\vec{a}$ were derived to be time reversible despite
the fact that the entropic argument clearly stipulates an arrow of time. More
generally, we showed that the probability of any \emph{continuous} path is
independent of the direction in which it is traversed. (Incidentally, if the
paths were not continuous but composed of small discrete steps then the
predictions would include tiny T violations.)

The second comment concerns the \textquotedblleft physical\textquotedblright%
\ origin of the position uncertainties, an important issue about which we have
remained silent. The fact that particle masses are a manifestation of these
uncertainties, $\sigma_{0}^{2}\propto1/m$, might be a clue. Among the various
approaches to quantum theory the version developed by Nelson, and known as
stochastic mechanics \cite{Nelson6685}, is particularly attractive because it
derives quantum theory from the hypothesis that particles in empty space are
subject to a peculiar Brownian motion characterized by position fluctuations
such that $\sigma^{2}\propto\hbar/m$. It is difficult to avoid the conclusion
that the uncertainties underlying entropic dynamics might be explained by
quantum effects. However, while this is a very tempting possibility, an even
more interesting and radical conjecture is that the explanatory arrow runs in
the opposite direction. The radical conjecture would be that the same entropic
dynamics that already explains mass, and interactions, and Newton's mechanics,
might also---and with no further assumptions---explain quantum mechanics as
well. Perhaps \emph{physics is nothing but inference} after all.

\end{document}